\pdfoutput=1
\documentclass[english]
{IEEEtran}
\usepackage{amsmath}
\usepackage{amssymb}
\usepackage{stmaryrd}
\usepackage{hyperref}
\usepackage{graphicx}
\usepackage{subfig}
\usepackage{physics}
\usepackage{todonotes}
\usepackage[T1]{fontenc}
\usepackage[utf8]{inputenc}

\title{Foundations of Intelligent Additive Manufacturing}
\author{
	\IEEEauthorblockN{K\'evin Garanger\IEEEauthorrefmark{1}, Eric Feron\IEEEauthorrefmark{1}, Pierre-Lo\"ic Garoche\IEEEauthorrefmark{2}, Julian J. Rimoli\IEEEauthorrefmark{1}, John D. Berrigan\IEEEauthorrefmark{3}, Martha Grover\IEEEauthorrefmark{4} and Kerianne Hobbs\IEEEauthorrefmark{1}}
	
	\IEEEauthorblockA{\small\IEEEauthorrefmark{1}School of Aerospace Engineering, Georgia Institute of Technology, 270 Ferst Dr. NW, Atlanta, GA 30032, USA}
	
	\IEEEauthorblockA{\small\IEEEauthorrefmark{2}ONERA - Centre de Toulouse, 2 Avenue Edouard Belin, 31000 Toulouse, France}
	
	\IEEEauthorblockA{\small\IEEEauthorrefmark{3}Materials and Manufacturing Directorate, Air Force Research Laboratory, Wright Patterson Air Force Base, OH 45433, USA}
	
	\IEEEauthorblockA{\small\IEEEauthorrefmark{4}School of Chemical \& Biomolecular Engineering, Georgia Institute of Technology, 311 Ferst Dr. NW, Atlanta, GA 30032, USA}
}
\date{}

\begin{document}
\maketitle

\begin{abstract}
	
  During the last decade, additive manufacturing has become increasingly popular
  for rapid prototyping, but has remained relatively marginal beyond the scope of
  prototyping when it comes to applications with tight tolerance specifications, such as in aerospace. Despite a strong desire to supplant many aerospace structures with printed builds, additive manufacturing has largely remained limited to prototyping, tooling, fixtures, and non-critical components. There are numerous fundamental challenges inherent to additive processing to be addressed before this promise is realized. One ubiquitous challenge across all AM motifs is to develop processing-property relationships through precise, \textit{in situ} monitoring coupled with formal methods and feedback control. We suggest a significant component of this vision is a set of semantic layers within 3D printing files relevant to the desired material specifications. This semantic layer provides the feedback laws of the control system, which then evaluates the component during processing and intelligently evolves the build parameters within boundaries defined by semantic specifications. This evaluation and correction loop
  requires on-the-fly coupling of finite element analysis and topology
  optimization. The required parameters  for this analysis are all extracted from the semantic layer and can be
  modified \textit{in situ} to satisfy
  the global specifications. Therefore, the representation of what is printed changes
  during the printing process to compensate for eventual imprecision or drift arising
  during the manufacturing process.
\end{abstract}

\section{Introduction}

During the design of a part, the choice of the geometry and of the material
depends on a set of specifications determined from the expected use of
the part. This choice is supposed to ensure that if a part is manufactured
according to the chosen design, all the specified requirements will be
satisfied. 
Each AM process has fundamental materials and processing limitations that affect ultimate performance of the build. The
certification of additive manufacturing (AM) processes is challenging because the large number of factors governing such
processes makes them difficult to precisely predict and reproduce. Often, the performance of a build can vary significantly due to the large number of processing variables combined with the prolonged, layer-by-layer nature of AM, which provides ample opportunity for random defects to occur. This inherent performance variability limits AM to niche applications where high performance can be sacrificed in favor of rapid production and customization while industries requiring stringent
specifications thus have tended to scant additive manufacturing for production and still mostly privilege traditional manufacturing
methods for critical components.

Improved performance of builds from AM tools would have significant impact to the manufacturing base -- particularly for medical, military, electronics and aerospace applications and when producing small batches of critical parts or highly customizable objects with strict specification, such as spacecrafts or customizable cars parts.

Since the certification of processes essentially consists of proving their
ability to produce builds respecting the given specifications, we believe that it
is crucial to include the information of the specifications in the manufacturing
process. Those specifications are a valuable knowledge which can be used in a
closed-loop control as a set of constraints and targets, both concepts stemming
from classical control theory.

Semantic annotations of AM files are a convenient way we adopted to specify the
expected performance of a build. With semantic annotations, most types of
specifications can be expressed and verified numerically though simulations
before, during and after the manufacturing process. Our approach extensively
relies on topology optimization through the use of finite element methods (FEM)
since they can be applied on arbitrary domains for a wide range of problems like
structural or thermal analysis, making them particularly suited for the online
analysis of any kind of 3D printed components.

\begin{figure*}[t]  
	\hspace{0mm}
	\begin{center}
		\includegraphics[width=14cm]{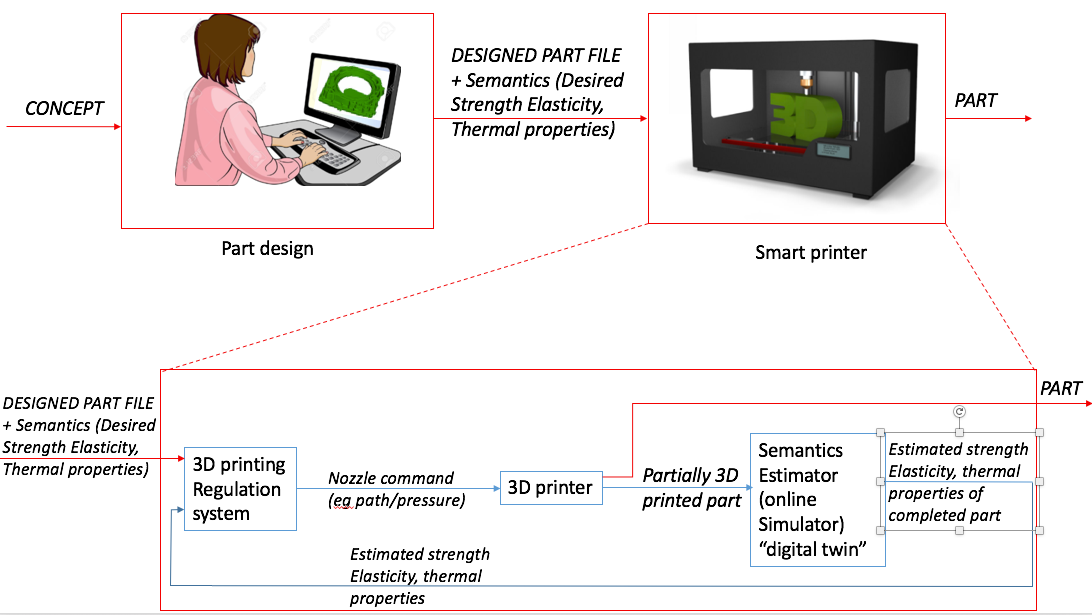}
	\end{center}
	\caption{Integrated 3D printing process}
	\label{process}
\end{figure*}

Our article proceeds as follows: \hyperref[semantics]{Section \ref*{semantics}}
presents the concept of semantics for AM files. \hyperref[inversion]{Section \ref*{inversion}} introduces the idea
of adapting the build local parameters on the fly while
\hyperref[control]{Section \ref*{control}} details the way to achieve the
necessary control precision to accomplish the desired topology. Finally, we
present conclusions and implications for further research.
The overall approach is sketched in Fig.~\ref{process}.

\section{Related works}\label{soa}

\subsection{Semantics in additive manufacturing}
A need exists within the community to develop the semantics of AM print files particularly when compared to the evolution of 
other sectors such as database management~\cite{Decker1999}, computational
finance~\cite{Zimmermann:2004:SGW:1028664.1028772} and reactive
software~\cite{bookchap_springer}. Only intents to annotate 3D shapes stored in
a database through concepts expressed by an ontology were made to ease
the retrieval process~\cite{attene2009characterization}. Another example of
semantic annotations for 3D shapes is in the frame of computer vision to extract
relevant object features~\cite{leifman2005semantic} or represent a robot
environment~\cite{rusu2010semantic}. However, none of these efforts closely
relate to the real-time manufacturing context described in this paper, whose concept is to using formal methods to allow a manufacturing process to evolve on the fly within the specifications of the semantic annotations.

\subsection{Closed-loop control in additive manufacturing}

Today, feedback control is, in general, a well-accepted discipline whose impact on 
the proper operation of complex systems can be beneficial. In some instances, 
systems could not operate without feedback control: Such is the case of nuclear plants and
advanced fighter aircraft~\cite{FPE:86}.

To improve the quality of AM produced builds, it is now commonly recognized that
one of the next steps is to create closed-loop and adaptive control systems with
feedback control
capabilities~\cite{bourell2009brief,frazier2014metal,huang2015additive}. Specific
cases of closed-loop and adaptive controls for AM have been developed. For
instance, adaptive and closed-loop control systems have been implemented in gas metal arc welding (GMAW) and powder bed fusion processes to control the deposition height based on a vision sensor~\cite{xiong2014adaptive} and to control the deposition
temperature of the Ti-6Al-4V alloy by altering the tool paths. While these demonstrations are the first steps towards real-time feedback and control, they are unable to account for varying materials properties or deviations from intended specifications. These geometric and performance based specifications form the foundational sets of targets and constraints that are utilized by classical control theory.
The recent publication~\cite{DebRoy2016} shows, however, the strong interest of the materials science community to address these new opportunities.

\section{Expressing the semantics of additive manufacturing files}\label{semantics}

The validation of a manufactured object requires non-destructive methods to evaluate and predict the overall performance of the build. This evaluation is a complex blend of inter-related parameters such as microstructure, density, electronic conductivity, geometry, void density, inter-layer adhesion, elastic modulus shear modulus, and anisotropy. Many of these relationships are inter-related, print-process dependent, and still unknown to the materials science community.

The validation of software, computer files or, more generally, a cyber-physical system (CPS) where
computer files are critical, is difficult. Because of the almost infinite set of
possible behaviors or, in general, outcomes from computer files, testing is
extremely expensive and can hardly reach a sufficient coverage of system
behaviors. For example compare a 1950's plant that manufactures nails and a 3D
printer. While the nail plant may have high capacity, its mechanical behavior is
very limited in comparison to the simplest 3D printer.

In computer science the use of formal methods addresses this issue by
representing the software as a mathematical object and analyzing it
exhaustively. The key notion here is the concept of program semantics. This
notion of semantics and formal analyses can be further applied to CPS. A CPS
combines a physical device, aka the plant, with a computer-based system,
typically a feedback controller. When designing the software controller the
plant model is characterized and used to setup the software parameters. This
closed loop system, combining the mathematical model of the plant and the formal
representation of the code through its semantics, can be analyzed in an exhaustive fashion.

We consider this approach for 3D printers which are special instances of CPS. In order to do so, a formal representation of the part which is printed is to be included to characterize accurately the system and its performance.

\subsection{Semantics of programs}

Let us develop this notion of semantics in the context of programs and see how
to apply it to the specific case of a 3D printer. Semantics describes the
meaning of a program. Consider for example a simple program computing the
Euclidian norm of a vector: it takes as input a vector \texttt{v} and returns a
scalar. The semantics of this program $\mathtt{norm}$ can be described by
different means: denotational semantics, operational semantics, or axiomatic
semantics. Denotational semantics associates a denotation, ie. a mathematical
function, to the program. It maps inputs to outputs without providing low level
details on the steps of the computation. In our case, it will be defined using
the mathematical definition of the Euclidian norm:

\[
\llbracket \mathtt{norm}
\rrbracket_{\mathrm{den}} \triangleq \lambda v. \sum_{i\in [0,sz(v)-1]} v(i)^2
\]
where $sz(v)$ returns the size of vector $v$.

The operational semantics are typically more detailed: they express the behavior
as a sequence of \textit{operations}. Steps of computation could be defined
through various means. One can, for example, rely on a control flow graph
representation describing sequences of imperative statements and guards
evaluation between program points, ie. states $\in \Sigma$. A state
$s \in \Sigma$ is associated to an environment mapping variables $v$ to values
$\nu$: $s:[v \mapsto \nu]$. The operational semantics are then described by a
transition system
$(\mathtt{init, trans, accept}) \in \Sigma \times \wp(\Sigma \times Statements
\times \Sigma) \times \wp(\Sigma)$
where \texttt{init} denotes the initial state, \texttt{trans} the list of
possible transitions from state to state and associated statements, and
\texttt{accept} the list of final states.

The operational semantics of our example program can be formalized as:
\[\llbracket \mathtt{norm} \rrbracket_{\mathrm{op}} \triangleq (s_1, T, \{s_3\})\]
where $\Sigma = \{s_1, s_2, s_3 \}$ \\and $T = \left \{ 
\begin{array}{l}
(s_1\;,\; \{r:= 0.0;\; i := 0;\};\;,\; s_2)\\
(s_2\;,\; \{i < sz(v);\; r:= r + v[i] * v[i];\}\;,\; s_2)\\
(s_2\;,\; \{i >= sz(v);\}\;,\; s_3)
\end{array}\right.$.
\bigskip

The last way to describe the behavior of the program is its axiomatic
semantics. While both denotational and operational semantics capture precisely
the semantics of the program and its complete behavior, axiomatic semantics
could be less precise and act as a projection semantics. A program or function
is fitted by a pre- and post-condition, also known as assume-guarantee
contract. It was first introduced by \textsc{Hoare} in
1969~\cite{DBLP:journals/cacm/Hoare69} as both a proof framework ``\emph{used to
	prove the correctness of the program [...] to place great reliance on the
	result of the program, and predict their properties with a confidence limited
	only by the reliability of the electronics}'' as well as the
``\emph{ultimately definitive specification of meaning of the program}''.

In this example, these axiomatic semantics can loosely specify the positivity of
the result or be more specific and impose it to be a sum of square:
\[
\begin{array}{l}
\llbracket \mathtt{norm} \rrbracket^1_{\mathrm{ax}} \triangleq (true, \mathtt{norm}(v) \geq 0)\\
\llbracket \mathtt{norm} \rrbracket^2_{\mathrm{ax}} \triangleq (true,
\mathtt{norm}(v) = \sum_i v[i]^2)
\end{array}
\]
A piece of code can be associated with multiple axiomatic contracts since they do
not necessarily capture its complete behavior but only part of it. 

In computer science the use of formal methods addresses this issue by
representing the software as a mathematical object and analyzing it
exhaustively. The key notion here is the concept of program semantics. This
notion of semantics and formal analyses can be further applied to CPS. A CPS
combines a physical device, aka the plant, with a computer-based system,
typically a feedback controller. When designing the software controller the
plant model is characterized and used to set up the software parameters. This
closed loop system combining the mathematical model of the plant and the formal
representation of the code could be analyzed in an exhaustive fashion.

\subsection{Semantics of additive manufacturing files}

\begin{figure*}
	\begin{center}
		\subfloat[Attaching semantics to models: certified autocoders]{%
			
			\includegraphics[width=.65\textwidth]{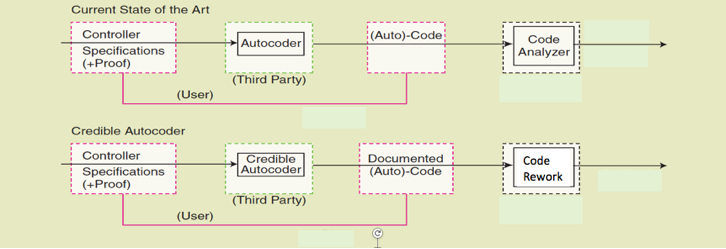}
		}~~%
		\begin{minipage}[b]{.3\linewidth}
			\small In~\cite{bookchap_springer,Wan:15} the classical compilation scheme of Simulink models in C
			code is enriched by expliciting model properties as synchronous
			observers. These additional annotations express both safety properties,
			eg. specification of redundancy patterns, and control ones, eg. open- or
			closed-loop stability of the controller thanks to the use of a Lyapunov
			function. These semantics elements are render explicit and are used at code
			level to validate the properties on the code artifact.
		\end{minipage}
		\\
		\begin{minipage}[b]{.3\linewidth}
			\small The proposed approach follows a similar process: the intended behavior
			of the 3D object model is made explicit and attached to the model: this
			characterizes its semantics. Along the printing process this semantics
			annotation is used in real-time to assess the evalidity of the printed object,
			providing a certified printing process. This axiomatic semantics describe the
			object in a \emph{projective} manner. These observers can be used as runtime
			verifiers.
		\end{minipage}~~%
		\subfloat[Attaching semantics to 3D models: certified printers]{%
			\includegraphics[width=.65\textwidth]{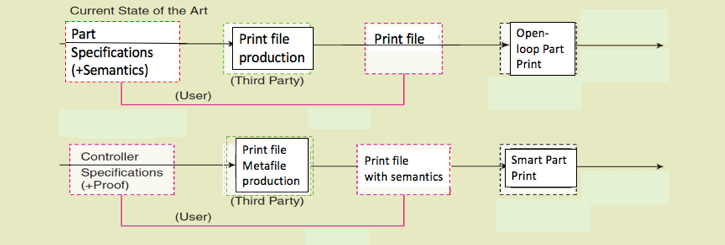}
		}
	\end{center}
	\caption{Semantics-based processes}
	\label{fig:sem_process}
\end{figure*}

To verify the performance of a manufactured build, the computer file of the build should come with enough information to assert the correctness of the
specifications of a printed build. We include directly this information in the
file as semantic annotations or as a metadata file when the file format does not support annotations, following the example of the semantics of
programming languages. The comparison between the two approaches is outlined in Fig.~\ref{fig:sem_process}.

In AM, the program is the object to be evaluated, ie. printed. Axiomatic
semantics can be used to declare \emph{projectively} the expected behavior or
meaning of that object. This behavior can describe properties of the
manufactured object: its thermal resistance, compressive strength or
stiffness/rigidity.

\subsubsection{Categorizing semantic properties of AM files}

There are two main categories of semantic properties for AM files. The first category is the one of properties that can be verified directly with the information of the file, that is its boundary representation, while the second category contains the other properties, those verifiable only under some assumptions about the material used for manufacturing.
The first category may for example contain the volume of an object, its number of faces, or some aerodynamic properties. Some properties that can be included in the second category are the mass of an object, its stiffness, or its conductivity.
Therefore when including properties of the second category to the semantic of the file, verifying these properties requires to specify the assumptions made on the material properties in the semantic annotations.
And the verification of the semantic properties is a crucial task and actually the main stake of introducing this new layer of metadata to AM files. Moreover, the role of the semantic annotations is not only to describe the properties that are expected given a set of material properties, but to help \emph{prove} them by decomposing the problem in easily verifiable assumptions, and to help \emph{reach} them by providing a guideline for the control system of the 3D printer during the manufacturing process.

The properties of the latter category, that is the ones that require some assumptions on the printing material, can also be subdivided in two subcategories: \emph{global} and \emph{local} properties. Global properties are properties describing an overall behavior of the printed object. For example, the mass of a build, its electrical resistance between two defined points, and its stress-strain analysis are all global properties. Although such properties rely on a local evaluation of the material parameters over the domain of the build, they do not make sense locally.
Local properties are properties that can be expressed locally considering the material parameters. Since we are interested in proving a set of specifications on the whole build, local properties can be assembled at the scale of the build following different rules. For instance, one can decide that a local property should be true over the a specific domain of the build. The temperature of a build under specific thermal flux and temperature boundary conditions is a local property.

\subsubsection{Formal definition of AM files properties}

During the manufacturing process, the inversion problem described in Section \ref{inversion}, which is
performed on the fly, relies on a finite element analysis (FEA) of a volumetric mesh
of the build. Even though other analysis methods can be considered in the long-run, such as multi-physics and multi-scale modeling to accurately understand properties. Finite element methods can solve most nontrivial properties of manufactured builds and can be easily described by the semantic annotations of an AM file. The same way the semantic of programs can help proving them by decomposing an algorithm in small natively provable statements, a volumetric mesh corresponds to a decomposition of a build model that is used to prove relevant complex properties.
A volumetric mesh is defined as a set of vertices or nodes with incidence relations that
form edges, faces, and elements. The geometry of the mesh specifies the positions
of the vertices and other eventual geometric characteristics of the vertices,
whereas the topology of the mesh describes the incidence relations between its
vertices, edges, faces, and elements.
Therefore the metadata included in a particular AM file is supposed to contain the whole information
necessary to reconstruct the volumetric mesh used for the FEA. This mesh has
been determined prior to the manufacturing in order to provide a sufficient
accuracy of the the simulations within a reasonable time. Many techniques exist
to generate a volumetric mesh suitable for a FEA from a surface mesh.
For instance, TetGen~\cite{si2006quality}, which is a software for tetrahedral mesh generation, can be used to generate a good quality mesh from an AM file. Most of the time, the surface mesh described in an AM file is not directly usable for a FEA since the Delaunay tetrahedralization of its vertices may not even contain all the edges and faces of the surface mesh. In that case a constrained Delaunay tetrahedralization can be performed. A constrained Delaunay tetrahedralization is a tetrahedralization of a set of vertices which contains some predefined edges while being the closest of a Delaunay tetrahedralization of those vertices. See~\cite{shewchuk2008general} for more details. 

FEM can provide the approximate solution of a partial differential equation
involving two quantities over the domain of the mesh. With FEM, local linear
equations approximating the behavior of the part at each node are assembled in a
linear equation of the form
\begin{equation}\label{fem_eq}
F = KU
\end{equation}
where $F$ and $U$ are the two vectors representing the quantities of the PDE at
each node and $K$ a symmetric positive definite matrix whose coefficients are
directly related to the properties of the material at the nodes of the mesh.

Usually, for a specific node, one of the involved quantities is known and the
other is unknown. The unknown values are obtained by solving \ref*{fem_eq} .
Since AM files originally only contain a polygonal surface mesh of the part, the
first information to be added to the polygon mesh is the remaining set of
vertices used to form a suitable volumetric mesh for a FEA. This would result in
a significant but manageable increase of the file size. This volumetric mesh
forms the basis for the FEA. Most of the time a Delaunay triangulation of the
vertices is sufficient. But it is also possible to consider a different
triangulation of the vertices or a tessellation that involves other polyhedra
than tetrahedra. In that case the semantic annotations contain the topological
information of the mesh (edges, faces, elements).

A volumetric mesh is then obtained on a domain $\Omega$ whose boundary
$\partial\Omega$ is exactly the surface mesh described by the AM file.
To each vertex is added all the necessary metadata to describe the constraints that have to be met and the parameters that are involved in the verification of those constraints.

Therefore, we do not exactly consider properties of AM files, but rather properties of tuples $(M, f)$, where $M$ is a volumetric mesh and $f$ a function which associates to each vertex of $M$ a set of parameters with their respective ranges. Formally, a property is a set of such tuples closed under mesh isomorphism. We say that $(M, f)$ verifies the property $P$ if $(M, f) \in P$.
For a printed part associated to a volumetric mesh $M$, the function $f$ is modified during the manufacturing process. Indeed, the vertices belonging to the domain of the part that has already been printed are associated new values with respect to $f$ and the observation of the process. While before printing $f$ associates to vertices expected parameter ranges, it then gives real values or ranges.
We are aware that a printed part can be correctly represented by different couples $(M, f)$ that do not verify the same properties. We do not address this problem though since it is more a question about the reliability of FEM in general, and an accurate mesh is assumed to be chosen.

\subsection{Example of a stress-strain analysis of a shaft}

To illustrate the use of semantic annotations in a specific problem, let's consider a polyhedral mesh describing a shaft of a single material which is supposed to support a mass $M$ while in upright position on a horizontal flat surface.

A stress-strain analysis of the shaft can be done with the direct
stiffness method, which is a particular case of FEM.
The Hooke's law gives the local relation between the strain and the stress. It allows us to establish for each vertex $i$ the force contribution $f_i^e$ from the
element $e$ applied to $i$ with respect to the displacements of the vertices $u_j$
\[f_i^e = \sum_j k_{ij}^e u_j\] 
Since the sum of the internal forces should equal the external forces at each
node,
\[f_i^{ext} = \sum_e f_i^e = \sum_{e,j}k_{ij}^eu_j = \sum_j K_{ij}u_j,\]
this gives the global stiffness equation
\begin{equation}\label{stiffness equation}
F^{ext} = KU
\end{equation}
For each vertex $i$, either $f_i^{ext}$ or $u_i$ is known. In this specific case, for
each vertex $i$ of the lower face of the shaft, $u_i$ = 0 and the other
displacements are unknown. For every vertex $i$ which is not in one of the two
horizontal faces, $f_i^{ext} = 0$. For the vertices of the upper face, the
forces are known and their sum is equal to the force applied by the mass $M$ on the
shaft. The forces at the lower face are unknown.
In this problem, the local stiffness matrices $k_{ij}^e$ and the
forces $f_i^{ext}$ do not have pre-determined fixed values but rather lie
respectively in some sets $\mathcal{K}_{ij}^e$ and $\mathcal{F}_i$, and the goal
is to ensure that the unknown $u_i$ stay within some defined sets
$\mathcal{U}_i$. \\
\textit{Remark}: since the interpolation functions for a specific element $e$
are linear and because we use a tetrahedral mesh, if $i_1,i_2,i_3,i_4$ are the nodes
of $e$, for every point $x$ of $e$, its displacement $u(x)$ lies within the
convex hull of $\bigcup_{i=1}^4\mathcal{U}_i$. \\
For each vertex $i$, the information of both sets $\mathcal{F}_i$ and
$\mathcal{U}_i$ are added in the semantic annotations. One a them represents the condition we are given and the other the goal we want to meet.  As well, each topological element $e$ comes with the geometrical information of its nodes, that is their positions, and its Young's modulus and Poisson's ratio $E \in \mathcal{E}_e$ and $\nu \in \mathcal{N}_e$, two parameters necessary to calculate the matrices $k_{ij}^e$.

\subsection{Semantics metadata attached to the CAD source file}

Similarly to the annotation languages used in formal verification to associate
formal specification to code artifact, eg.~\cite{acsl},  an annotation language for CAD models can be developed. This additional information may be redundant but addresses both means to recompute the information such as  additional mesh data and computed results such as local forces or stiffness matrices. Annotations for CAD files denote both global and local properties, and metadata can also include argumentation means, ie. computation strategies support the (re)computation of global properties from local ones, simplifying the verification of specifications.
A credible autocoding environment can be developed from CAD files to AM files to obtain the annotated AM files from the CAD files.

\section{Adapting the build local properties during the manufacturing process}\label{inversion}

While the final build is printed, the monitoring of the manufacturing process can evaluate the performance of the build with regard to the semantics of the AM file and what has already been printed. The semantic annotations are used to produce a prediction of the build after termination of the process and to assess if the build specifications are respected or not. If there is a chance that the build will not satisfy the requirements described by the semantic of its file, the printing of the rest of the build can be adapted to meet the final constraints by changing the expected values of the material parameters present in the semantic annotations, and by changing the control system variables adequately.
In most cases, a AM print file and its semantics will exhibit a prescribed geometry constant material properties over the domain of the build. However, the printing process will introduce inherent microscopic defects affecting the material property. With a careful monitoring of these defects and an adapted analysis, if the altered material properties can be estimated on the fly, with micro-mechanical models for example, the control parameters of the printing process can accordingly be modified to obtain different material parameters for the remainder of the build, thus compensating for the initial discrepancy in the expected material properties. In that case, even if the build is designed with already defined constant material properties, the final build is actually printed with varying local properties, optimized during the printing the process.
Following the example of the shaft, adapting the material parameter means determining a new matrix $K$ that contains the information of the known stiffness coefficients $k^e_{ij}$ of the vertices belonging to the domain of what has already been printed, and that yields a solution to the stiffness equation (\ref{fem_eq}) respecting the specifications. Of course, when resolving this problem, the manufacturing constraints must be taken into account too. Whereas in a stiffness equation such as (\ref{fem_eq}), $K$ is usually given and one seeks to find the unique solution of the equation, we do not consider that the matrix $K$ is fixed. Our goal is rather to be able to modify the matrix $K$, so that the solutions of the stiffness equation are satisfying. This is why we later refer to this step as the \emph{inversion problem}.

A way to solve the inversion problem while incorporating all the manufacturing constraints is to perform a multi-scale topology optimization of the rest of the manufactured build to optimize the material properties.

\subsection{Online topology optimization}

\begin{figure*}[t]  
	\hspace{0mm}
	\begin{center}
		\includegraphics[width=14cm]{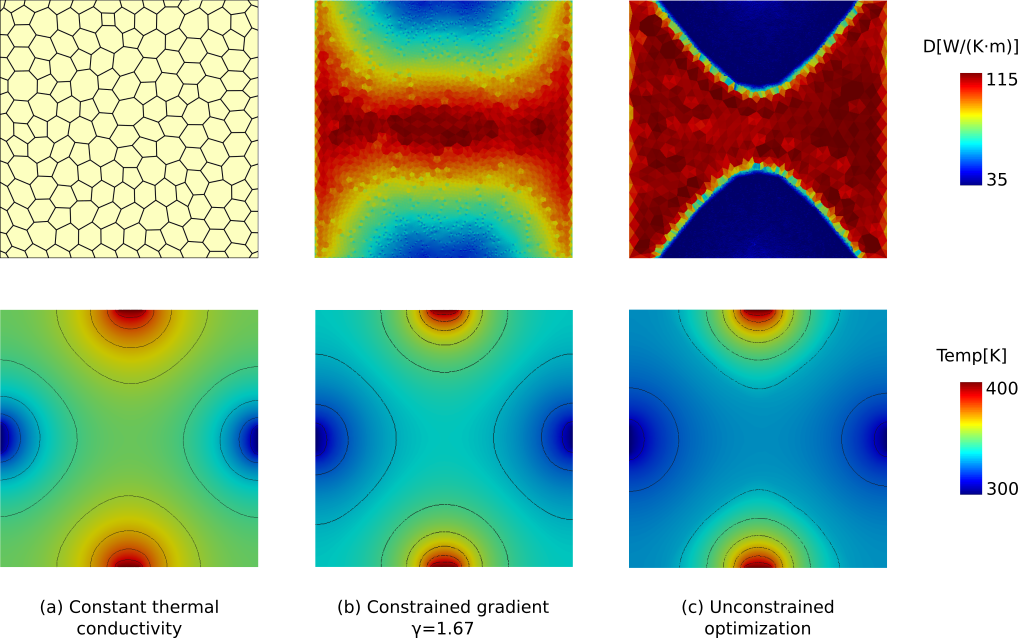}
	\end{center}
	\caption{Example of multi-scale topology optimization in which the microstructure of the component is optimized pointwise for optimal thermal transport \cite{bouquet2016exploiting}.}
	\label{thermal}
\end{figure*}

Topology optimization is generally adopted to optimize the macroscopic (continuum scale) topology of an object given a set of constraints, loads, and boundary conditions. It is usually a minimization problem which is solved during the design step of the object with various methods approximating a solution like gradient descent or genetic algorithms. In contrast to traditional approaches to topology optimization, recent work \cite{bouquet2016exploiting} combines topology optimization techniques with advanced microstructural models \cite{bouquet2015length} to optimize the microstructural distribution within a component rather than its macroscopic topology. This multi-scale optimization approach is particularly well suited to 3D printing applications, as this manufacturing process inherently produces microstructures that affect the local material properties of the component, e.g., through the introduction of voids or preferred directions, and hence its performance. One of the main advantages of topology optimization methods is that many various constraints can be expressed in the system to optimize. For instance, if the manufacturing process requires a local real-valued continuous parameter like the thermal conductivity $D(x)$ to be $\gamma$-Lispchitz, we can add the constraint $|D(x)-D(y)| \leq \gamma||x-y||$ to the optimization problem to be solved. Figure \ref{thermal} is an example of the multi-scale topology optimization on the thermal conductivity of a build with some temperature boundary constraints to minimize the average temperature. In the case of AM, we can proceed to the same optimization step but also extend it to the manufacturing process. In the case where the mesh is fixed, the parameters to be optimized are the values of the key parameters at the nodes belonging to the domain of the object that is not yet printed. The same parameters but for the nodes corresponding to the part of the build already printed become fixed values that are involved in the optimization process but not optimized since they cannot be changed anymore. This online topology optimization process allows us to take into account the defects that arise during the manufacturing process and that modify the optimization problem.
For the computational efficiency of the online topology optimization, one can imagine methods that do not perform \textit{de novo} the whole minimization step at each step, but reuse the previous results to estimate a new minimum during the manufacturing of the build. For instance, let's consider that we have a real-valued smooth function $F$ whose domain is an open of $\mathbb{R}_n$ and that represents the function to be optimized with respect to $x\in  \mathbb{R}_n$, $x$ being the set of local material parameters that we can control. Let say that $F$ is minimized for $x_0 = (y_0, z_0)$ with $y_0$ representing the first $p$ coordinates of $x_0, p<n$. Let's assume that the Hessian matrix of $F$ at this point is positive definite. The printing process evolves with the objective of reaching $x_0$, and the first $p$ coordinates of $x$ are now fixed to the value $y_0 + \var y$ because the printing of the build is not perfectly accurate. Then, instead of recalculating the minimum of $F(y_0 + \var y, z)$ with respect to $z$, one can write $F(y_0 + \var y, z) = F(y_0 + \var y, z_0 + \var z)$ and make the approximation 
\begin{align*}
F(y_0 + \var y, z_0 + \var z) \approx &\ F(y_0, z_0) \\
&+ F_y(y_0, z_0)\var y + F_z(y_0, z_0)\var z \\
&+ \frac12 \var y^T F_{yy}(y_0, z_0)\var y \\
&+ \frac12 \var z^T F_{zz}(y_0, z_0) \var z \\
&+ \var z^T F_{yz}(y_0, z_0)\var y \\
= &\ F(y_0, z_0) \\
&+ \frac12 \var y^T F_{yy}(y_0, z_0)\var y \\
&+ \frac12 \var z^T F_{zz}(y_0, z_0) \var z \\
&+\var z^T F_{yz}(y_0, z_0)\var y \\
\end{align*}
Assuming that $F_{zz}(y_0, z_0)$ is definite positive, such quantity can be minimized by taking
\[
\var z = F_{zz}(y_0, z_0)^{-1}F_{yz}(y_0, z_0)\var y
\]
This estimation allows us to save some computational resources, especially since minimization problem can be hard to solve in high-dimensional continuous spaces.

To obtain the desired macroscopic properties, the topology optimization acts on the microscopic parameters that can be modified during the printing process. Those are not always directly controllable but rather influenced by the way the closed-loop control system is parametrized. By modifying the closed-loop control of a 3D printer, we can aim to create specific microstructures to obtain the wanted values for local parameters in the topology optimization.
This approach is developed in ~\cite{bouquet2016exploiting} where a microstructural
optimization is performed, enabling the computation of multiscale optimization by tailoring
microstructure to obtain desired macroscopic properties.
Topology optimization can notably be used to achieve a desired stiffness by
minimizing the final weight, for example with Michell
structures ~\cite{michell1904lviii}. ~\cite{zheng2005structural} studies the topology optimization of structures subject to design-dependent loads while ~\cite{stromberg2010efficient} studies tradeoff curves for topology optimization under design constraints. ~\cite{doubrovski2011optimal,brackett2011topology,murr2012metal,zhang2015efficient} suggest methods more specific to AM that we can adapt for online optimization by manufacturing specific open-cellular structures with pre-selected stiffness.

\subsection{Material parameters monitoring}
In order to be able to incorporate the local parameters of the material of what has been printed to the topology optimization of the manufactured build, one must be able
to evaluate these parameters with a sufficiently good precision in real-time. To monitor the
manufacturing process of a build one may think of many different possibilities,
like the use of a profilometer, of a scale measuring the mass of the build like
in~\cite{hu2003sensing} or of a thermal camera like
in~\cite{griffith1999understanding}. \cite{tapia2014review,mani2015measurement}
give several examples of monitoring processes and ways to correlate raw
measurements to mechanical properties. With the profile history, thermal history
or any other pertinent data of the build that can be measured in real-time, the relevant
properties can be reconstituted with various statistical methods. One can
consider the extensive use of Bayesian filtering and inference to estimate the properties
which are described in the semantic layer of the 3D manufacturing files and are
supposed to be feedback-controlled.
Having an accurate estimation of the parameters of the printing process with little delay is necessary for the topology optimization, but also to use a closed-loop control based on the optimized parameters for each element of the mesh. The control aspect of the 3D printing is detailed in the next section.

\section{Control}\label{control}

\subsection{Relation between control variables and topology}

During the optimization step of inversion problem, the parameters used are the ones which are the most relevant to the analysis of the AM build with respect to the specified constraints of the semantics. These are variables independent of the 3D printer that is used for the manufacturing and that are only specific to the printed build. However, he 3D printer controls a set of different parameters that can be totally different from an AM process to another and even from a printer to another. In order to be able to bring the semantic parameters to the values determined by the solution of the inversion problem, the impact of the 3D printer control variable on those semantic parameters should be understood. This understanding of the incidence relationships between all these variables and parameters can be described as a learning process. This learning process relies on a modeling step and on specific experiments to refine the model.

\subsubsection{Modeling of the printing process}

In order to be able to modify the printed object topology as desired, the relationships between the control variables and the material properties must be known. A modeling step is necessary to predict the behavior of the 3D printer in the actual environment of the manufacturing process.
\cite{hu2003modelling,toyserkani20043,costa2005rapid,bontha2006thermal}
give several models especially concerned with thermal processes during specific
additive manufacturing. Such models can be used during the feedback control,
parametrized and given more or less confidence when performing a Markov
estimation of the state of a manufactured build.
Examples of modeled properties of the material are its microstructure and the presence of anisotropic tensiles. Because of the
layer-by-layer deposition, AM methods create by nature anisotropies in the
printed material \cite{gao2015status}. These anisotropic properties have been
studied for different types of materials, such as the polymer
ABS~\cite{ahn2002anisotropic} or the Ti-6Al-4V titanium
alloy~\cite{baufeld2011wire,carroll2015anisotropic}. Both materials are
frequently used in AM, and the latter is particularly relevant 
since it is one of the most common alloys in aerospace industry, a sector
directly concerned by low-volume production, and is widely used for biomedical
applications like implants and prostheses, two kinds of devices that benefit
from efficient mass-customization processes. Moreover, 3D manufactured builds often suffer from end product surface roughness, and their micro structural characteristics
are strongly affected by their thermal history,
especially for processes like Laser-based additive manufacturing (LBAM), during
which a printed build is subject to high temperature gradients and important
heating and cooling rates influence the microstructure of the
material~\cite{griffith1999understanding,kobryn2001mechanical,bontha2006thermal,%
	zheng2008thermal,bontha2009effects,thijs2010study,gaytan2010comparison,%
	antonysamy2012microstructure,shamsaei2015overview}. Thermal history is also
known to create
micro-hardnesses~\cite{mazumder1997direct,griffith1999understanding,costa2005rapid,%
	el2008phase,shamsaei2015overview} and residual
stress~\cite{rangaswamy2005residual,liu2011microstructure,shamsaei2015overview}
in the printed build, degrading it's performance. Modeling those phenomena is a necessary step to achieve the desired optimized topology.

\subsubsection{Refining the model with experiments on test builds}
As every model used to describe a problem, the one we can obtain with the ideas explained in the previous paragraph cannot be perfect, and the more parameters and relationships we intend to exhibit, the more like it is that the model will be inaccurate. Therefore experiments can be performed on test builds in order to refine the model and have a better understanding of the printing process. This is the equivalent of the exploration step necessary in many machine learning problems dealing with the well-known dilemma of the \emph{exploration-exploitation trade-off}. Such test builds are cantilever builds, they allow us to calibrate 3D printers by determining inherent stress during the manufacturing of simple builds.
For a specific AM process or a specific 3D printer, one can intend to define relevant printing processes of test builds to accomplish this exploration step until it is satisfying enough to switch to the exploitation step. This exploitation step being the the manufacturing of the build.
One can also imagine a context for performing exploration steps during the printing of the final build to take in account more precisely the specificities of the printed build and the present printing environment.

\section{Conclusion}
The objective of this work introduced in this paper is to create a semantic environment for AM files in order to draw a link between the geometrical description of the part, the printing material properties, and the part specifications. With such a semantic environment, a formal description of the part can be provided along with the geometrical description of the part. This environment provides semantic annotations to the volumetric mesh describing the part, so that a topology optimization of the part associated to a closed-loop control is performed during the manufacturing process to thwart the eventual unpredicted printing defects that can arise. This concept could only be effective if the AM processes were understood well enough. For many of them, however, this is not the currently the case, so that the type of information to be included in semantic annotations for AM print files remains unclear. In a long-term vision, the definition of a formal environment to determine appropriate control laws should be general enough to adapt to the progress that can be made in term of AM processes modeling and understanding.

\section{Acknowledgments}
This work was supported by grants CPS 1446758 and CPS 1544332 from the National Science Foundation.
Prof. Rimoli would like to acknowledge the support by the National Science Foundation Division of Civil, Mechanical, and Manufacturing Innovation under Grant No. CMMI-1454104.

\bibliographystyle{IEEEtran}
\bibliography{document}


\end{document}